\documentclass[conference]{IEEEtran}
\IEEEoverridecommandlockouts
\usepackage{cite}
\usepackage{amsmath,amssymb,amsfonts}
\usepackage{algorithmic}
\usepackage{graphicx}
\usepackage{graphics}
\usepackage{algorithm}
\usepackage{algorithmic}
\usepackage[algo2e,linesnumbered,ruled]{algorithm2e}
\usepackage{multirow}
\usepackage{array}
\usepackage{textcomp}
\usepackage{xcolor}
\def\BibTeX{{\rm B\kern-.05em{\sc i\kern-.025em b}\kern-.08em
    T\kern-.1667em\lower.7ex\hbox{E}\kern-.125emX}}

\newtheorem{definition}{Definition}

\SetKwProg{Fn}{Function}{}{end}

\SetAlgoSkip{}

\newcolumntype{M}[1]{>{\centering\arraybackslash}m{#1}}

\DeclareSymbolFont{symbolsC}{U}{txsyc}{m}{n}
\DeclareMathSymbol{\notniFromTxfonts}{\mathrel}{symbolsC}{61}

\makeatletter
\newcommand*\titleheader[1]{\gdef\@titleheader{#1}}
\AtBeginDocument{%
	\let\st@red@title\@title
	\def\@title{%
		\vskip-1.1em\bgroup\normalfont\footnotesize\centering\@titleheader\par\egroup\vskip.4em\st@red@title\vskip-.1em}
}
\makeatother

\title{Probabilistic Skyline Query Processing over Uncertain Data Streams in Edge Computing Environments\\
	\thanks{This research is partially supported by Ministry of Science and Technology, Taiwan under the Grant No. MOST 107-2221-E-027-099-MY2 and MOST 109-2634-F-009-018- through Pervasive Artificial Intelligence Research (PAIR) Labs, Taiwan.}
}

\titleheader{This is the accepted version of the work. The final version will be published in 2020 IEEE GLOBECOM, December 7-11, Taipei, Taiwan}

\author{\IEEEauthorblockN{Chuan-Chi Lai\IEEEauthorrefmark{1}, Yan-Lin Chen\IEEEauthorrefmark{2}, Chuan-Ming Liu\IEEEauthorrefmark{2}, and Li-Chun Wang\IEEEauthorrefmark{1}}
	\IEEEauthorblockA{\IEEEauthorrefmark{1}Deptartment of Electrical and Computer Engineering, National Chiao Tung University, Taiwan\\
		\IEEEauthorrefmark{2}Deptartment of Computer Science and Information Engineering, National Taipei University of Technology, Taiwan\protect\\
		Email: cmliu@ntut.edu.tw
	}%
}

\begin{document}

\maketitle

\begin{abstract}
	With the advancement of technology, the data generated in our lives is getting faster and faster, and the amount of data that various applications need to process becomes extremely huge. Therefore, we need to put more effort into analyzing data and extracting valuable information. Cloud computing used to be a good technology to solve a large number of data analysis problems. However, in the era of the popularity of the Internet of Things (IoT), transmitting sensing data back to the cloud for centralized data analysis will consume a lot of wireless communication and network transmission costs. To solve the above problems, edge computing has become a promising solution. In this paper, we propose a new algorithm for processing probabilistic skyline queries over uncertain data streams in an edge computing environment. We use the concept of a second skyline set to filter data that is unlikely to be the result of the skyline. Besides, the edge server only sends the information needed to update the global analysis results on the cloud server, which will greatly reduce the amount of data transmitted over the network. The results show that our proposed method not only reduces the response time by more than 50\% compared with the brute force method on two-dimensional data but also maintains the leading processing speed on high-dimensional data.
\end{abstract}


\section{Introduction}\label{sec:introduction}
\begin{figure}[!t]
	\centering
	\includegraphics[width=.475\textwidth]{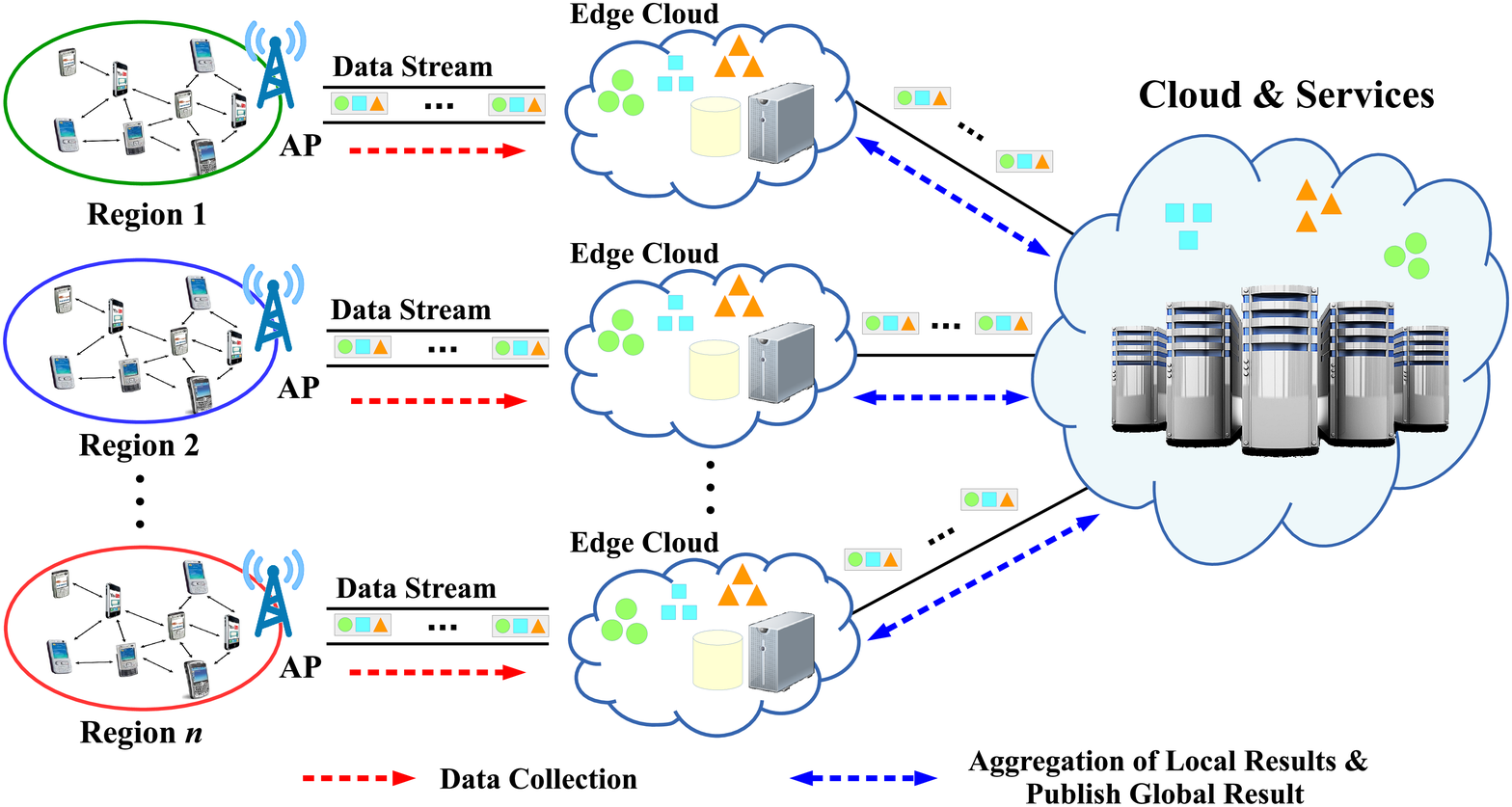}
	\vspace{-5pt}
	\caption{The considered edge computing environment.}
	\label{fig:system_model}
	\vspace{-15pt}
\end{figure}

As the Internet of Things (IoT) generates more and more data, edge computing has become a promising computing model that can handle big data streams to provide rapid response to meet the low latency requirements of emerging applications~\cite{7488250}~\cite{8731646}. Unlike the cloud computing model that uses large computing server clusters to deal with big data problems, the edge computing model allocates more computing resources on edge servers. Such a way can effectively reduce the response time of processing big data streams and quickly answer user queries received. As a result, this prompted us to propose a query processing method for stream computing services based on an edge computing environment. The considered edge computing environment is depicted in Fig.~\ref{fig:system_model}.

In this work, we consider a kind of query, \emph{Skyline}. Skyline query is a spatial data processing technology for multiple criteria decision making (also known as multi-objective optimization or Pareto optimization) problems. Skyline is also called as \emph{Pareto frontier} in Pareto optimization.
Although skyline query has been under discussion for many years, most of the skyline query processing methods~\cite{10.1145/1061318.1061320}~\cite{10.1145/1559845.1559898} are designed in centralized computing environments. 
Papadias \textit{et al.}~\cite{10.1145/1061318.1061320} proposed an approach, \emph{Branch-and-Bound Skyline} (BBS), based on the best-first nearest neighbor search algorithm~\cite{Hjaltason:1999:DBS:320248.320255}.
Zhang \textit{et al.}~\cite{10.1145/1559845.1559898} proposed a hybrid framework including filter and sampling steps for minimizing the communication cost of monitoring frequent skyline query in client-server computing architecture.
Compared to the centralized solutions, some research~\cite{10.1007/s10115-009-0269-0}~\cite{KOH2017114} discuss skyline query in distributed computing environments. Sun \textit{et al.}~\cite{10.1007/s10115-009-0269-0} proposed an tree-based indexing approach, \emph{GridSky}, for processing skyline queries over distributed certain data streams in a master-slave computing cluster. The slave computing nodes calculate their local skylines. Then, the master computing node incrementally updates the final skyline after receiving all the local skylines. GridSky can help both slave and master computing nodes to prune the irreverent data, and thus the communications cost between slave and master nodes can be reduced. Koh \textit{et al.}~\cite{KOH2017114} proposed a parallel skyline processing framework, \emph{MR-Sketch}, based on MapReduce system. They applied different data partition policies to mapper step and reducer step in MR-Sketch, and then discussed the performance of skyline query processing with different partition policies. 

The above approaches only consider certain data. Processing uncertain data is much more complex than the certain data and the system requires more computation costs. Due to the uncertainty of data attributes, the skyline result may have many combinations and the set of these combinations is called as \emph{probabilistic skyline}.
Zhang \textit{et al.}~\cite{ZHANG20131212} proposed a centralized computing framework for deriving the skyline over sliding windows on uncertain data elements against probability thresholds in real-time. Gavagsaz~\cite{Gavagsaz2020} proposed a parallel skyline processing framework based on MapReduce system for processing probabilistic skyline queries, but this work did not support streaming computing for real-time monitoring.

In this work, we hence consider how to process and monitor the skyline query over uncertain data streams collected from emerging applications in the forthcoming IoT era. The contributions of this work are listed as follows.
\begin{itemize}
	\item To the best of our knowledge, very few research discussed the real-time skyline query processing over uncertain data streams in edge computing environments.
	\item We also proposed a new method, \emph{Probabilistic Second Skyline Update} (PSSU), for effectively pruning irrelevant information so as to improve the performance of processing skyline query over uncertain data streams.
	\item According to the simulation result, the proposed method significantly improves the performance of processing skyline queries in terms of response time and average transmission cost.
\end{itemize}

The rest of paper is organized as follows. 
Section~\ref{sec:problem} presents the preliminary and problem statement.
Section~\ref{sec:proposed_approach} explains the proposed approach with algorithms and examples in detail. In Section~\ref{sec:simulation}, we present the simulation results and validate the performance of the proposed method in various situations. Finally, the conclusion remarks of this work are given in Section~\ref{sec:conclusion}.



\section{Preliminary and Problem Statement}
\label{sec:problem}

\subsection{Preliminary}
Data with uncertainty are called uncertain data. There are mainly 3 different kinds of uncertain data [1], fuzzy model, evidence-oriented model, and probabilistic model. Furthermore, the probabilistic model can be divided into continuous model and discrete model. The continuous model represents data with a probabilistic density function (PDF). The PDF of data object $u_i$ in a continuous uncertain data model is represented as $\text{pdf}(u_i)$, and $\text{pdf}(u_i)=\int_{x\in u_i}\text{pdf}(x)dx=1$.
In our work, we consider uncertain data with the discrete probabilistic model and the discrete probabilistic data model can be defined as follows.

\begin{definition}[\textbf{Discrete Probabilistic Data Model}]
	\label{def:DPDM}
	Given an uncertain data object $u_i$ which is composed of $j$ instances, denote as $u_i=\{u_{i,1},u_{i,2},\dots,u_{i,j}\}$. Each instance has a probability of occurrence $\text{Pr}(u_{i,j})$. The occurrence probability of uncertain data object $u_i$ is the sum of the occurrence probabilities of all instances and it can be denoted as $\text{Pr}(u_{i})=\sum_{u_{i,j}\in u_{i},\forall j}u_{i,j} \leq 1$.
\end{definition} 

The data comes into our system is represent as a $n$-sphere according to the data dimension. That is, given a center point $c$ of uncertain data object $u_i$, and the corresponding radius $r$, all instance of $u_i$ will locate inside the $n$-sphere. In another word, the Euclidean distance between $c$ and any instance of $u_i$ will not exceed $r$. 
Take Table~\ref{table:2d_uncertain_data_set} for example. It shows a set of uncertain data objects. Each data object has 3 instances and each instance contains 2 attributes.
%

\begin{table}[!t]
	\centering
	\footnotesize
	\caption{An Example of a 2D Uncertain Data Set}
	\label{table:2d_uncertain_data_set} 
	\begin{tabular}{|c|c|c|c|}
		\hline
		\textbf{Object}            & \textbf{Instance} & \textbf{Probability} & \textbf{Attributes} \\ \hline
		\multirow{3}{*}{$u_1$} &   $u_{1,1}$    &    0.4    &    [28,47]   \\ \cline{2-4} 
		&   $u_{1,2}$    &    0.1    &   [27,45]    \\ \cline{2-4} 
		&   $u_{1,3}$   &     0.5     &   [25,48]   \\ \hline
		\multirow{3}{*}{$u_2$} &   $u_{2,1}$    &    0.1      &    [9,35]    \\ \cline{2-4} 
		&   $u_{2,2}$   &     0.2     &   [9,38]   \\ \cline{2-4} 
		&   $u_{2,3}$   &     0.7    &   [10,33]   \\ \hline
		\multirow{3}{*}{$u_3$} &   $u_{3,1}$    &     0.5     &     [24,92]    \\ \cline{2-4} 
		&   $u_{3,3}$   &     0.3     &   [22,91]     \\ \cline{2-4} 
		&   $u_{3,3}$   &     0.2     &   [22,88]   \\ \hline
	\end{tabular}
\end{table}

\begin{table}[!t]
	\centering
	\footnotesize
	\caption{A Sliding Window Example}
	\label{table:sliding_window} 
	\begin{tabular}{|c|c|c|}
		\hline
		\textbf{Time} & \textbf{Sliding Window} & \textbf{Size} \\ \hline
		1	&  $W_1=\{u_1\}$  &  $|W_1|=1$    \\ \hline
		2	&  $W_2=\{u_1,u_2\}$  &  $|W_2|=2$    \\ \hline
		3	&  $W_3=\{u_1,u_2,u_3\}$  &  $|W_3|=3$    \\ \hline
		4	&  $W_4=\{u_2,u_3,u_4\}$  &  $|W_4|=3$    \\ \hline
		5	&  $W_5=\{u_3,u_4,u_5\}$  &  $|W_5|=3$    \\ \hline
	\end{tabular}
	\vspace{-10pt}
\end{table}

In a data stream system, data will continuously come in. Data usually comes with a timestamp and expires after a period of time. Expired data provide no information to further analysis and may provide incorrect information. As a result, when analyzing data in a data flow system, outdated data must be filtered out. Thus, the analysis result without incorrect or irrelevant information can provide valuable insights to users. Sliding window is a technique to tackle data streams. Because of the characteristic of data stream that outdated data provide no information, sliding window is a suitable technique to be used. There are two types of sliding windows, time-based and count-based. In the time-based sliding window, data will be removed from the sliding window if the time it stays in sliding window exceed a given time period. In our research, we use count-based sliding window to implement our proposed approach. The count-based sliding window is defined as follows.

\begin{definition}[\textbf{Count-Based Sliding Window}]
	\label{def:rnn}
	A sliding window at time $t$ is denote as $W_t$. One sliding window will have a maximum size $n$, denote as $|W|=n$. The size of sliding window at time $t$ is denote as $|W_t|$. In any time, $|W_t|$ will not exceed the maximum size $n$. That is $|W_t|\leq n, \forall t$.
\end{definition} 

Assume $u_4$ comes at $t=1$, $u_5$ comes at $t=2$, and so on. The maximum size of sliding window $W$ is 3. That is $|W|=3$. Table~\ref{table:sliding_window} gives a example to show the change of sliding window from $t=1$ to $t=5$.

To search the probabilistic skyline, the system needs to calculate the dominant relations between different uncertain objects and instances. The instance-level dominant relations can be defined as follows.

\begin{definition}[\textbf{Instance-Level Dominance}]
	\label{def:instance_dominant}
	Given two instances of two different data objects, $p_x$ and $p_y$. Object $p_x$ dominates $p_y$, denote as $p_x\prec p_y$, if and only if all the attribute of $p_x$ is less or equal to $p_y$, and exists one attribute that $p_x$ is less than $p_y$. That is, the probability of the instance-level dominance for $p_x$ with respect to $p_y$ is derived by	
	\vspace{-5pt}
	
	\footnotesize
	\begin{align*}\label{eq:instance_dominant_probability}
		\text{Pr}(p_x\prec p_y)=&\begin{cases}
			\text{Pr}(p_x)\cdot \text{Pr}(p_y), & \text{if $(p_x.attr(i)\leq p_y.attr(i), \forall i)$}\\
			& \wedge (p_x.attr(j)<p_y.attr(j), \exists j).\\
			0, & \text{otherwise}.
		\end{cases}
	\end{align*}
\end{definition} 

For example, given two 3D data, $d_1=[11,5,7]$ and $d_2=[15,5,10]$. We can say that $d_1$ dominates $d_2$ which is denoted as $d_1\prec d_2$.

Since a data object may has multiple instances, some instances of an object dominates some instances of another object, but some does not. In addition, each instance has its own probability of existence. As a result, the object-level dominate relationship will be a dominance probability which is the sum of instance-level dominance probabilities. The definition of object-level dominance probability is presented below.

\begin{definition}[\textbf{Object-Level Dominance}]
	\label{def:object_dominant}
	Given two uncertain data objects $u_x$ and $u_y$, where $x\neq y$. The dominating probability of $u_x\prec u_y$ is
	\begin{equation*}\label{eq:obj_dominant_probability}
		\text{Pr}(u_x\prec u_y)=\sum_{p_{x,i}\in u_{x},p_{y,j}\in u_{y},\forall i,j}\text{Pr}(p_{x,i}\prec p_{x,j}).
	\end{equation*}
\end{definition}

\subsection{Problem Statement}
Consider an edge computing environment with uncertain data sources. There are $n$ edge computing nodes (ECNs) $E_1, E_2,\dots, E_n$ with adequate computing resources and a main server $S$. All the data comes into ECNs are uncertain data streams. Our goal is to find a global skyline set on server $S$ according to the information provide by ECNs nodes. Since our research focus on edge computing environment, it is necessary to reduce the amount of data transmission from ECNs to the main server as much as possible. Also, the algorithm should keep the accuracy of skyline set. The time to calculate probabilistic skyline set is also an important factor since data streams are time sensitive, the execution time has to be minimized.

\section{Proposed Query Processing Framework}
\label{sec:proposed_approach}
In this section, we will introduce the design of the proposed query processing framework in detail. Table~\ref{table:notations} shows the frequently used notations in the proposed algorithm.

\begin{table}[!ht]
	\centering
	\footnotesize
	\caption{Frequently Used Notations}
	\label{table:notations} 
	\begin{tabular}{|c|c|}
		\hline
		\textbf{Notation} & \textbf{Meaning} \\ \hline
		$W$ & Sliding window \\ \hline
		$u_i$ & Uncertain data object $i$ \\ \hline
		$u_{i,j}$ & Instance $j$ of uncertain object $u_i$  \\ \hline
		$ESK_{i,1}$ &  The 1st skyline candidate set of ECN $E_i$  \\ \hline
		$ESK_{i,2}$ & The 2nd skyline candidate set of ECN $E_i$  \\ \hline
		$SK_1$ & The 1st skyline candidate set of main server $S$ \\ \hline
		$SK_2$ & The 2nd skyline candidate set of main server $S$ \\ \hline
	\end{tabular}
	\vspace{-5pt}
\end{table}

\subsection{Probabilistic Second Skyline Update Algorithm (PSSU)}
In order to reduce the time to calculate the skyline set and minimize the amount of transmitted data over the networks, we proposed the Probabilistic Second Skyline Update algorithm (PSSU). In PSSU algorithm, each ECN is responsible for calculating the 1st local skyline set, $ESK_{i,1}$ and the 2nd local skyline set, $ESK_{i,2}$. Once the required update occurs at an ECN. The ECN will send an update message to main server. The main server is in charge of maintaining the global skyline candidate sets, $SK_1$ and $SK_2$, and performs any updates according to the information provided by ECNs.

PSSU uses R-Tree~\cite{10.1145/602259.602266} as the indexing method. By utilizing the advantages of R-Tree, data can retrieve fast and accurately without accessing a lot of irrelevant data points. Such a way effectively reduces execution time and thus satisfies our requirement of fast update.
After obtaining the skyline set from sliding window, the second skyline candidate set is the skyline set of those in sliding window which are not in the first skyline set. The formal definition of second skyline candidate set is described as follows.
\begin{definition}[\textbf{The Second Skyline Candidate Set}]
	\label{def:2nd_skyline_set}
	Given a sliding window $W$ and the corresponding skyline set $SK=Skyline(W)$. The second skyline set, marked as $SK_2$, will be $SK_2=Skyline(W-SK)$.
\end{definition} 

We use the second skyline candidate set as a pruning method. The second skyline candidate set reduces the amount of lookup data required when updating is needed because data that is not in the skyline set or the second skyline set will never become a skyline object. Therefore, it is not necessary to consider those irrelevant data when performing the update step.

\subsection{Data Indexing}
\begin{figure}[!ht]
	\centering
	\includegraphics[width=.3\textwidth]{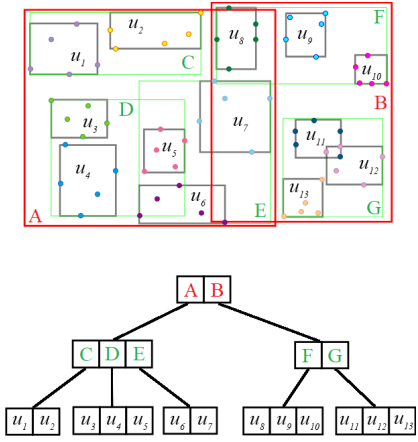}
	\caption{An example of R-tree.}
	\label{fig:ex:r_tree}
	\vspace{-15pt}
\end{figure}

\normalsize
As mentioned before, PSSU uses R-Tree for data indexing. Since the data objects entering the system have uncertain instances, each dimension of the data has maximum and minimum values. PSSU stores the \emph{minimum bounded rectangle} (MBR) of a data object as an index entry (or a leaf node of the R-tree).
By storing the MBR of each uncertain data object in the index, they can be treated as certain data when pruning the irrelevant information. This data indexing technique help the system reduces time of calculating the dominating probability without accessing the unnecessary data objects. Fig.~\ref{fig:ex:r_tree} shows an example of R-tree. Assume there are 13 two-dimensional data objects with 5 instances stored in the index. The rectangles represent the MBRs of objects stored in R-Tree.

\subsection{The Tasks of an Edge Computing Node}
Each ECN $E_i$ is in charge of calculating local skyline. After obtaining two candidate sets $ESK_{i,1}$ and $ESK_{i,2}$, $E_i$ will compare old skyline sets with the new one to check if there is any change in local skyline. If any updates are required, $E_i$ will send a update message to the server $S$. The update message of $E_i$ contains the following information: (1) the new data in $ESK_{i,1}$, (2) the new data in $ESK_{i,2}$, and (3) the outdated data of $E_i$. The server can update the global skyline set according to the received messages from ECNs. Basically, there are two main tasks on each ECN, \emph{Receive} and \emph{Update}. In the followings, we will explain the procedures of these tasks in detail.

We denote the PSSU procedure on the ECN as EPSSU. Algorithm~\ref{alg:EPSSU} describes the operations of an ECN. First of all, edge will save the state of $ESK_{i,1}$ and $ESK_{i,2}$ before accepting the incoming data. After updating the local skyline, edge can compare current status with previous one and send update information to server.
When a new data stream $s_i$ comes into ECN $E_i$, $E_i$ will check if the sliding window is full. If the sliding window, $W$, is full, $E_i$ will remove the outdated data from the sliding window $W$ and then add the incoming new data objects in the input data stream $s_i$ to the sliding window, $W$. These operations are in the function, $\mathsf{ReceiveData}(s_i,W)$. ECN $E_i$ calls this function at the line~\ref{alg:EPSSU:3} of Algorithm~\ref{alg:EPSSU}. Algorithm~\ref{alg:EPSSU:receive} shows the detailed procedure of $\mathsf{ReceiveData}(s_i,W)$.

After ECN $E_i$ obtained the outdated data, $E_i$ updates the local skyline set at line~\ref{alg:EPSSU:4} of Algorithm~\ref{alg:EPSSU}. The detailed operations of updating skyline set are in the function, $\mathsf{UpdateSkyline}(ESK_{i,1},ESK_{i,2},s_i,outdated\_data)$, which is presented in Algorithm~\ref{alg:EPSSU:update}. Since removing data from $ESK_{i,2}$ does not affect the result, there is no need to take any caution in this step. Also, outdated data need to be removed from $ESK_{i,1}$. When removing data from $ESK_{i,1}$, it is necessary to check if there exist any data in $ESK_{i,2}$ that are dominated by the data being removed, and move those data from $ESK_{i,2}$ to $ESK_{i,1}$ because those data might become one of the skyline data. After all thet outdated data are removed, the new data will be moved into $ESK_{i,1}$ for the further updates. Next, the ECN will compute the new $ESK_{i,1}$ and $ESK_{i,2}$. Note that some data from $ESK_{i,2}$ are moved to $ESK_{i,1}$ and new data are added directly into $ESK_{i,1}$. Those data are not verified if it belongs to $ESK_{i,1}$ yet. As a result, checking if there exist dominant relationships between data in $ESK_{i,1}$ is required. If any data objects in $ESK_{i,1}$ are dominated by other data objects in $ESK_{i,1}$, the dominated data objects need to be moved the to $ESK_{i,2}$ since these data objects do not satisfy the skyline property. The similar work flow also applies on $ESK_{i,2}$. The difference is that if there exist any data in $ESK_{i,2}$ dominated by other in $ESK_{i,2}$, the edge would drop the data being dominated from $ESK_{i,2}$ directly. Finally, the ECN will remove all the outdated data collected previously.

\begin{algorithm2e}[!t]
	\footnotesize
	\SetAlgoLined
	\KwIn{Uncertain data stream $s_i$, Sliding window $W_i$}
	\tcc{save current status of skyline set 1}
	$oldESK_{i,1}\leftarrow \mathsf{getSkyline1}()$\;
	\tcc{save current status of skyline set 2}
	$oldESK_{i,2}\leftarrow \mathsf{getSkyline2}()$\;
	\tcc{receive new data and get outdated data}
	$outdated\_data\leftarrow\mathsf{ReceiveData}(s_i,W)$
	\label{alg:EPSSU:3}\;
	$\mathsf{UpdateSkyline}(oldESK_{i,1},oldESK_{i,2},s_i,outdated\_data)$\label{alg:EPSSU:4}\;	
	$ESK_{i,1}\leftarrow \mathsf{getSkyline1}()\setminus oldESK_{i,1}$\;
	$ESK_{i,2}\leftarrow \mathsf{getSkyline2}()\setminus oldESK_{i,2}$\;
	$\mathsf{SendResult}(outdated\_data,oldESK_{i,1},oldESK_{i,2})$\;
	\caption{The procedure of EPSSU on ECN $E_i$}
	\label{alg:EPSSU}	
\end{algorithm2e}
%

\begin{algorithm2e}[!t]
	\footnotesize
	\SetAlgoLined
	\KwIn{Uncertain data stream $s_i$, Sliding window $W_i$}
	\KwOut{Outdated data set $outdated\_data$}
	\If{$|W_i|\geq|W_{\max}|$}{
		$outdated\_data\leftarrow W.\mathsf{collectOutdatedData}()$\;
		$W.\mathsf{removeData}(outdated\_data)$					
	}
	$W.\mathsf{addData}(s_i)$\;
	\Return $outdated\_data$\;
	\caption{$\mathsf{ReceiveData}(s_i,W_i)$}
	\label{alg:EPSSU:receive}
\end{algorithm2e}
%

\begin{algorithm2e}[!t]
	\footnotesize
	\SetAlgoLined
	\KwIn{Two local skyline sets $ESK_{i,1}$ and $ESK_{i,2}$, Uncertain data stream $s_i$, Outdated Data Set $outdated\_data$}	
	\ForEach{data object $o$ in $outdated\_data$}{
		Remove data object $o$ from $ESK_{i,2}$\;
	}
	\ForEach{data object $o$ in $ESK_{i,1}$}{
		\If{$o.\mathsf{isOutdated}()$}{
			Move data object $o'$ into $ESK_{i,1}$ if $o\prec o', \forall o'\in ESK_{i,2}$\;
		}				
	}
	$ESK_{i,1}.\mathsf{append}(s_i)$\;
	\ForEach{data object $o$ in $ESK_{i,1}$}{
		\If{$o\prec o', \forall o, o'\in ESK_{i,1}$}{
			Move data object $o'$ into $ESK_{i,2}$\;
		}				
	}
	\ForEach{data object $o$ in $ESK_{i,2}$}{
		\If{$o\prec o', \forall o, o'\in ESK_{i,2}$}{
			Remove data object $o'$ from $ESK_{i,2}$\;
		}				
	}
	\tcc{remove all the outdated data in this ECN}	
	$\mathsf{ClearAllOutdatedData}()$\;
	\caption{\protect \\ $\mathsf{UpdateSkyline}(ESK_{i,1},ESK_{i,2},s_i,outdated\_data)$}
	\label{alg:EPSSU:update}
\end{algorithm2e}

\subsection{The Tasks of the Main Sever}
The main server takes the responsibility of maintaining the global skyline set. When the server receives update information from any ECN, it will start the update procedure. The following algorithms describe tasks of the server in detail.

We refer PSSU procedure on the server side as SPSSU. Algorithm~\ref{alg:SPSSU} shows the operations of SPSSU. When server receives information from edge, the update procedure begins. When the update procedure is finished, server will wait until another update information is received. After the server receives the message form ECN $E_i$, it will remove outdated data first. Then, for new data in $ESK_{i,1}$ and $ESK_{i,2}$, the server will check if those data are already in the server or not. If the received data is already in the sliding window $W_s$ on the server, it means that the data has been moved, so server needs to move the data to $ESK_{i,1}$ or $ESK_{i,2}$ according to dominant relationships between the received data and those are already stored in the sever. If the data is not in the server, server will need to store the received data to the sliding window $W_s$. The detailed operations are described in Algorithm~\ref{alg:SPSSU:reseive}.

The procedure of updating the global skyline, $\mathsf{UpdateGlobalSkyline}(W_s,SK_1,SK_2, outdated\_data)$, is presented in Algorithm~\ref{alg:SPSSU:update}
In fact, the procedure of update global skyline is almost identical to the procedure of updating local global skyline on each ECN. Thus, the detailed explanations of the algorithm would be skipped.

\begin{algorithm2e}[!t]
	\footnotesize
	\SetAlgoLined
	\KwIn{Uncertain data stream $s$, Sliding window $W_s$, Global skyline set $SK_1$, Global skyline candidate set $SK_2$}
	\While{true}{
		\If{$|s|>0$}{
			\tcc{call Algorithm~\ref{alg:SPSSU:reseive}}
			$outdated\_data\leftarrow\mathsf{ReceiveLocalUpdate}(s,W_s,SK_1,SK_2)$\;
			\tcc{call Algorithm~\ref{alg:SPSSU:update}}
			$\mathsf{UpdateGlobalSkyline}(W_s,SK_1,SK_2,outdated\_data)$\;
		}				
	}
	\caption{The procedure of SPSSU on server $S$}
	\label{alg:SPSSU}
\end{algorithm2e}

%
\begin{algorithm2e}[!t]
	\footnotesize
	\SetAlgoLined
	\KwIn{Uncertain data stream $s$, Sliding window $W_s$, Global skyline set $SK_1$, Global skyline candidate set $SK_2$}
	\KwOut{Outdated data set $outdated\_data$}
	Parse the receive data stream $s$ and then get the local outdated data set $outdated\_data$, the first local skyline candidate set $ESK_1$, and the second local skyline candidate set $ESK_2$\;	
	\ForEach{data object $o$ in $outdated\_data$}{
		Remove data object $o$ from $W_s$\;
	}
	\ForEach{data object $o$ in $ESK_1$}{
		\If{data object $o$ is not in $W_s$}{
			$W_s.\mathsf{add}(o)$\;	
			$SK_1.\mathsf{add}(o)$\;		
		}
		\ElseIf{data object $o$ is in $SK_2$}{			
			$SK_2.\mathsf{remove}(o)$\;
			$SK_1.\mathsf{add}(o)$\;
		}
	}
	\ForEach{data object $o$ in $ESK_2$}{
		\If{data object $o$ is not in $W_s$}{
			$W_s.\mathsf{add}(o)$\;	
			$SK_2.\mathsf{add}(o)$\;		
		}
		\ElseIf{data object $o$ is in $SK_1$}{			
			$SK_1.\mathsf{remove}(o)$\;
			$SK_2.\mathsf{add}(o)$\;
		}
	}
	\Return $outdated\_data$\;
	\caption{$\mathsf{ReceiveLocalUpdate}(s,W_s,SK_1,SK_2)$}
	\label{alg:SPSSU:reseive}
\end{algorithm2e}
%

\begin{algorithm2e}[!t]
	\footnotesize
	\SetAlgoLined
	\KwIn{Sliding window $W_s$, Global skyline set $SK_1$, Global skyline candidate set $SK_2$}
	\ForEach{data object $o$ in $outdated\_data$}{
		Remove data object $o$ from $SK_2$\;
	}	
	\ForEach{data object $o$ in $SK_1$}{
		\If{$o.\mathsf{isOutdated}()$}{
			Move data object $o'$ into $SK_1$ if $o\prec o', \forall o'\in SK_2$\;
		}				
	}
	\ForEach{data object $o$ in $SK_1$}{
		\If{$o\prec o', \forall o, o'\in SK_1$}{
			Move data object $o'$ into $SK_2$\;
		}				
	}
	\ForEach{data object $o$ in $SK_2$}{
		\If{$o\prec o', \forall o, o'\in SK_2$}{
			Remove data object $o'$ from $SK_2$\;
		}				
	}
	\caption{\protect\\
		$\mathsf{UpdateGlobalSkyline}(W_s,SK_1,SK_2, outdated\_data)$}
	\label{alg:SPSSU:update}
\end{algorithm2e}
%


\section{Simulation Results}
\label{sec:simulation}
We conduct several simulations to verify the performance of the proposed algorithm. We compare our approach with the brute force method in an edge computing environment. Both brute force method and PSSU calculate $ESK_{i,1}$ and $ESK_{i,2}$ on ECNs, but in different ways. When an ECN receives new data, the brute force approach will use all the data in the sliding window to re-calculate the first skyline candidate set and use all of the data that are not in the first skyline candidate set to calculate the second skyline candidate set. 

The simulations are executed on a computer with an Intel Core i7-4770 CPU and 16GB RAM. The operating system is Ubuntu 18.04. We use python 3.6.7 to implement our simulations. Table~\ref{table:simulation_settings} shows the default value of each parameter.

\begin{table}[!ht]
	\vspace{-8pt}
	\centering
	\footnotesize
	\caption{Detail Information}
	\label{table:simulation_settings} 
	\begin{tabular}{|l|c|}
		\hline
		\textbf{Parameter} & \textbf{Value} \\ \hline \hline
		The size of sliding window & 300 \\ \hline
		The number of edge computing nodes & 6 \\ \hline
		The number of data objects & 10,000 \\ \hline
		Data range & [0,1000] \\ \hline
		Data dimensionality & 2 \\ \hline
		Data radius & 5 \\ \hline
	\end{tabular}
	\vspace{-8pt}
\end{table}

\subsection{Impact of Data Dimensionality}
\begin{figure}[!t]
	\centering
	\includegraphics[width=.43\textwidth]{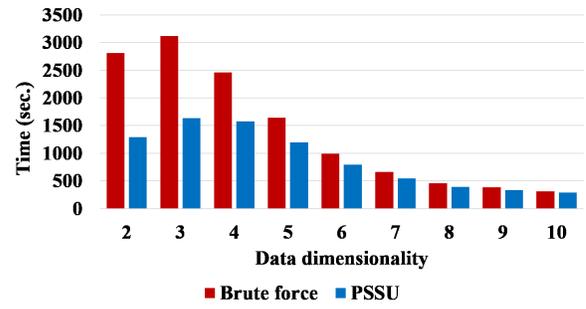}
	\vspace{-5pt}
	\caption{The response time of different approaches while varying the data dimensionality.}
	\label{fig:time_dimensionality}
\end{figure}

\begin{figure}[!t]
	\centering
	\includegraphics[width=.43\textwidth]{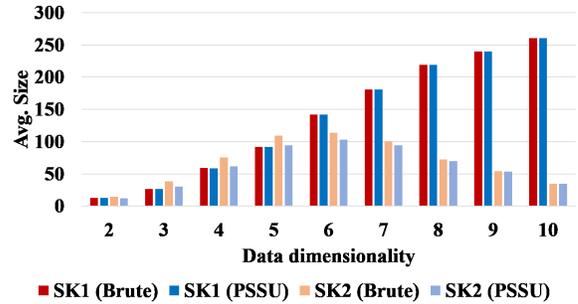}
	\vspace{-5pt}
	\caption{The average size of skyline set using different approaches while varying the data dimensionality.}
	\label{fig:skyline_size_dimensionality}
\end{figure}

In our simulation, we first observe how does data dimensionality affects the response time and the average size of skyline set. The result is shown in Fig.~\ref{fig:time_dimensionality}. It turns out that as data dimensionality increases, the response time of brute force approach and PSSU both decreases. However, the time difference between two approaches also decreases. Also, according to Fig.~\ref{fig:skyline_size_dimensionality}, as the data dimensionality increases, the average size of skyline set also increases. It turns out that almost all the data in the sliding window are also in either $ESK_{i,1}$ or $ESK_{i,2}$ when the data dimensionality is high. The reason is that data in sliding window are more likely to becomes skyline object as the data dimensionality increases. The number of pruned data objects using PSSU algorithm decreases. As a result, it can also be concluded that processing skyline queries on high-dimensional uncertain data is still a challenge.

\subsection{Impact on Transmission Cost}
\begin{figure}[!t]
	\centering
	\includegraphics[width=.43\textwidth]{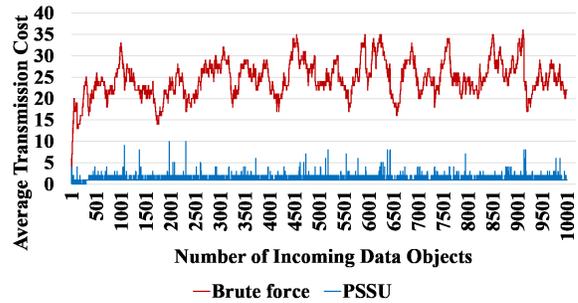}
	\vspace{-5pt}
	\caption{The average transmission cost of an ECN while varying the number of input data objects.}
	\label{fig:transmission_cost}
	\vspace{-5pt}
\end{figure}

In this simulation, we would like to know the average required transmission cost of information exchange between EDNs and the main server. In the consider edge computing environment, ECN $E_i$ generates a local skyline update information which contains three parts: (1) the first (local) skyline candidate set $ESK_{i,1}$, (2) the second (local) skyline candidate set $ESK_{i,2}$, and (3) the outdated data set. Hence, ECN $E_i$ using brute force approach needs to send a update message contains the all the data objects in the above three kinds of sets to the main server. However, compared to the brute force approach, ECN $E_i$ using EPSSU in Algorithm~\ref{alg:EPSSU} only need to send a update message which contains following three kinds of information: (1) new data objects in $ESK_{i,1}$, (2) new data objects in $ESK_{i,2}$, and (3) the outdated data set. In general. the size of an update message generated by EPSSU is much smaller than the one generated by brute force.
The simulation results of the above two approaches, PSSU and brute force, in terms of transmission cost, are shown in Fig.~\ref{fig:transmission_cost}. As we can see, PSSU costs much less transmission cost than the brute force approach. That is, with the help of PSSU, the amount of transmitted data for processing probabilistic skyline in the considered edge environment is effectively reduced.

\section{Conclusion}
\label{sec:conclusion}
In this work, we propose a new heuristic algorithm for processing probabilistic skyline queries on uncertain data streams in edge computing environments. The proposed algorithm, probabilistic second skyline update (PSSU), uses the concept of the second skyline set to prune irrelevant data, thereby reducing the response time. In addition, PSSU reduces the data transmission costs between edge computing nodes and the main server. The simulation results show that PSSU outperforms the brute force method. We also found that processing skyline query over high-dimensional uncertain data is still a big challenge. In the future, we are going to apply the proposed framework with some customized schemes and domain knowledge to the emerging multiple criteria decision making applications~\cite{9014210}~\cite{9177297}.

\bibliographystyle{IEEEtran}
\bibliography{reference}

\end{document}